\documentclass[a4paper,11pt]{article}

\usepackage{amsfonts,amssymb,amsmath,amsthm,dsfont,xfrac,xspace}
\usepackage{fullpage}
\usepackage{graphicx}
\usepackage{subfig}
\usepackage{float}
\usepackage{cite}
\usepackage{hyperref}
\usepackage{algorithmic}
\graphicspath{{./},{fig/}}
\makeatletter
\let\NAT@parse\undefined
\makeatother
\usepackage[sort&compress, numbers]{natbib}
\usepackage{hyperref}

\usepackage{verbatim}

\def\uu#1{\underline{\underline{#1}}}
\newcommand*{\rom}[1]{\expandafter\@slowromancap\romannumeral #1@}
\usepackage{amsmath}
\usepackage{resizegather}

\usepackage{mathtools,bbm}
\usepackage{cuted}
\usepackage{multicol}
\usepackage{stfloats}

\newsavebox{\ieeealgbox}

\renewcommand{\thelem}{\the \numexpr (\value{thm}+1) \relax.\arabic{lem}}

\title{Physical Layer Security in Frequency-Domain Time-Reversal SISO OFDM Communication} 

\author{Sidney~Golstein\thanks{S.~Golstein and J~. Sarrazin are with Labratoire d'Electronique et Electromagn\'etisme, Sorbonne Universit\'e (SU), 75005 Paris, France},\thanks{S.~Golstein, T.-H.~Nguyen, P.~De~Doncker, and F.~Horlin are with OPERA department, Universit\'e libre de Bruxelles (ULB), 1050 Brussels, Belgium. E-mail: sigolste@ulb.ac.be} , Trung-Hien~Nguyen \footnotemark[2] , Philippe~De~Doncker \footnotemark[2],\\ Fran\c cois~Horlin \footnotemark[2] , and Julien Sarrazin\footnotemark[1]  }

\begin{document}
\maketitle

\begin{abstract}
A frequency domain (FD) time-reversal (TR) precoder is proposed to perform physical layer security (PLS) in single-input single-output (SISO) system using orthogonal frequency-division multiplexing (OFDM). To maximize the secrecy of the communication, the design of an artificial noise (AN) signal well-suited to the proposed FD TR-based OFDM SISO system is derived. This new scheme guarantees the secrecy of a communication toward a legitimate user when the channel state information (CSI) of a potential eavesdropper is not known. In particular, we derive an AN signal that does not corrupt the data transmission to the legitimate receiver but degrades the decoding performance of the eavesdropper. A closed-form approximation of the AN energy to inject is defined in order to maximize the secrecy rate (SR) of the communication. Simulation results are presented to demonstrate the security performance of the proposed secure FD TR SISO OFDM system.\end{abstract}


\section{Introduction}
Due to their broadcast nature, wireless communications remain unsecured. With the  deployment of 5G as an heterogeneous network possibly involving different access technologies, physical layer security (PLS) has gained recent interests in order to secure wireless communications, \cite{PLS_litt1,PLS_litt2,PLS_litt3}. PLS classically takes benefit of the characteristics of wireless channels, such as multipath fading, to improve security of communications against potential eavesdroppers. A secure communication can exist as soon as the eavesdropper channel is degraded with respect to the legitimate user one, \cite{wyner}. This can be achieved by increasing the signal-to-interference-plus-noise ratio (SINR) at the intended position and decreasing the SINR at the unintended position if its channel state information (CSI) is known, and/or, by adding an artificial noise (AN) signal that lies in the null space of the legitimate receiver's channel. While many works implement these schemes using multiple antennas at the transmitter, only few ones intend to do so with single-input single-output (SISO) systems \cite{PLS_litt4,TR_FD_TD,TR_AN_2018_xu,TR_AN_2017_Li,TR_AN_2018_Li}.

In \cite{PLS_litt4}, a technique is proposed that combines a symbol waveform optimisation in time-domain (TD) to reach a desired SINR at the legitimate receiver and an AN injection using the remaining available power at the transmitter when eavesdropper's CSI is not known. Another approach to increase the SINR in SISO systems is time reversal (TR). This has the advantage to be implemented with a simple precoder at the transmitter. TR achieves a gain at the intended receiver position only, thereby naturally offering a possibility of secure communication, \cite{otges}. TR is achieved by up/downsampling the signal in the TD. It as been shown in  \cite{TR_FD_TD} that TR can be equivalently achieved in frequency domain (FD) by replicating and shifting the signal spectrum. FD implementation has the advantage to be easily performed using orthogonal frequency-division multiplexing (OFDM). To further enhance the secrecy, few works combine TD TR precoding with AN injection \cite{TR_AN_2018_xu,TR_AN_2017_Li,TR_AN_2018_Li}. In these works, the AN is added either on all the channel taps or on a set of selected taps. While the condition for AN generation is given, its derivation is however not detailed. Furthermore, the impact of the back-off rate (BOR), defined as the up/downsampling rate \cite{TR_bor}, has not been yet studied in the literature. 

An approach to establish secure communication using a FD TR precoder in SISO OFDM systems is proposed. An AN signal is designed to maximize the secrecy rate (SR) of the communication in presence of a passive eavesdropper whose CSI is supposed unknown. The proposed scheme uses only frequency diversity inherently present in multipath environments to achieve security. It can therefore be used in SISO systems and is then well-suited for resource-limited nodes such as encountered in Internet-Of-Things (IoT) for instance.  Indeed, MIMO capabilities require several antennas and as many transceivers and ADC/DAC, which might not fit into small-size sensors and could  be too power-consuming for such IoT scenarios. Furthermore, the OFDM implementation makes this approach compatible with LTE and 5G systems.

The reminder of this paper is organized as follows: the conventional FD TR-based OFDM system is presented in Section \ref{sec:model} as well as the way to design and inject the AN. In Section \ref{sec:perf}, a closed-form approximation of the amount of AN energy to be injected in order to maximize the SR is derived. Theoretical and numerical results are shown in Section \ref{sec:result}. Section \ref{sec:ccl} concludes the paper. \\
\textit{Notation:} the underlined upper-case letter denotes a column vector. Double-underlined upper-case letter corresponds to a matrix; $\uu{I}_N$ is the $N\times N$ identity matrix; $\left(.\right)^{-1}$, $\left( . \right)^*$, $\left(.\right)^H$ are respectively the inverse, the complex conjugate, and the Hermitian transpose operators; $\mathop{\mathbb{E}}$ is expectation operator; $x!$ is the factorial of a positive integer $x$.

\section{System Model}
\label{sec:model}
\subsection{Conventional FD TR SISO OFDM communication}
\label{subsec:traditional_FDTR}

\begin{figure}[t]
    \centering
    \centerline{\includegraphics[width = .53\textwidth]{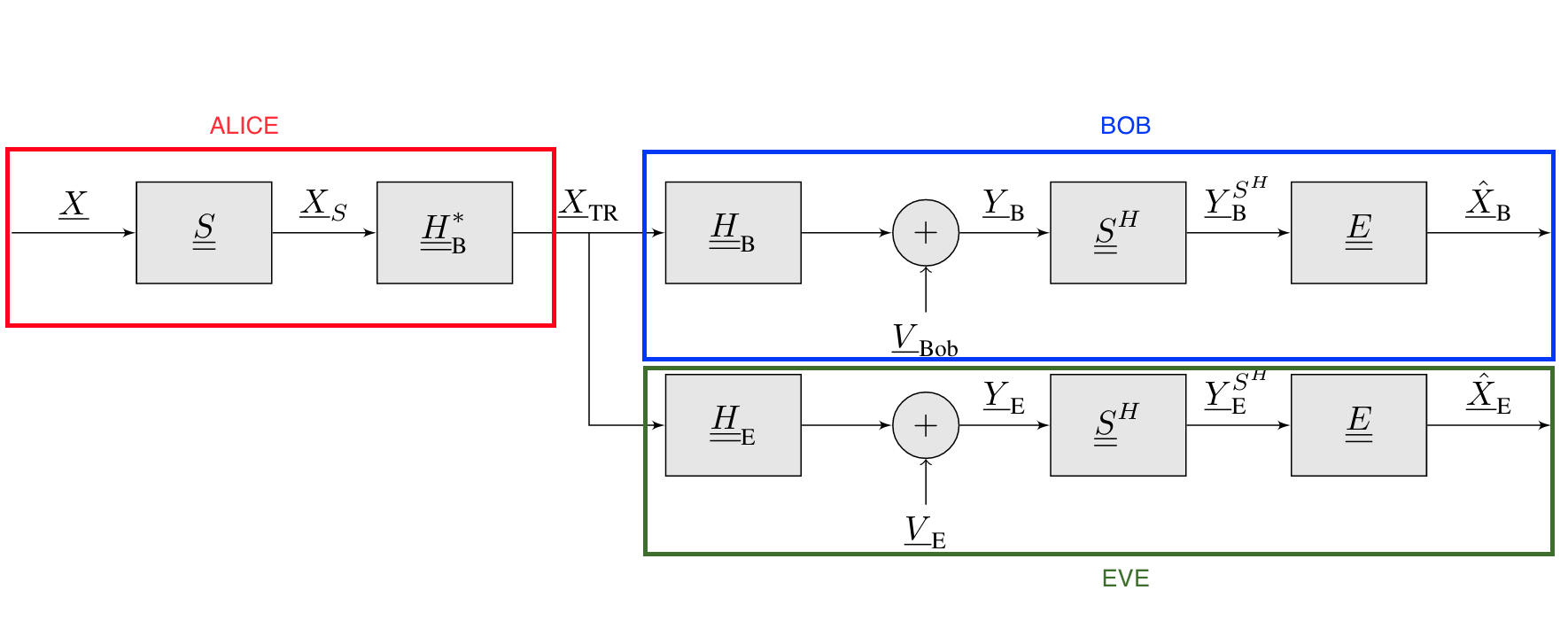}}
    \caption{Conventional FD TR SISO OFDM system}
    \label{fig:TR_FD_classical}
\end{figure}
The FD TR precoding scheme is illustrated in Fig. \ref{fig:TR_FD_classical}.  The communication is designed such that the data focuses at the legitimate receiver's position, i.e., at Bob. An eavesdropper, Eve, tries to intercept the data. We assume that the transmitter Alice does not have any information about Eve's CSI. The data is conveyed onto OFDM symbols with $Q$ subcarriers. Without loss of generality, we consider that only one OFDM block $\underline{X}$ is sent over the FD TR precoding SISO OFDM system. A data block $\underline{X}$ is composed of $N$ symbols $X_n$ (for $n = 0,..., N-1$, with $N\leq Q$). The symbol $X_n$ is assumed to be a zero-mean random variable with variance $\mathop{\mathbb{E}}\left[\left|X_n \right|^2\right] = \sigma_X^2 = 1$ (i.e., a normalized constellation is considered). The data block $\underline{X}$ is then spread with a factor $U = Q/N$, called back-of rate (BOR), via the matrix $\uu{S}$ of size $Q\times N$. The matrix $\uu{S}$ stacks $U$ times $N\times N$ diagonal matrices, with diagonal elements taken from the set $\{\pm1\}$ and being identically and independently distributed in order not to increase the peak-to-average-power ratio (PAPR) as suggested in \cite{papr}. 
This matrix is normalized by a factor $\sqrt{U}$ in order to have $\uu{S}^H \uu{S} = \uu{I}_N$:
\begin{equation}
\underline{\underline{S}} = \frac{1}{\sqrt{U}} \; . \;
   \begin{pmatrix}
    \pm 1 & 0 & \hdots & 0 \\
    0 & \pm 1 & \hdots & 0 \\
    \vdots & & \ddots & \vdots \\
    0 & 0 & \hdots & \pm 1 \\
     & \vdots& \vdots& \\
    \pm 1 & 0 & \hdots & 0 \\
    0 & \pm 1 & \hdots & 0 \\
    \vdots & & \ddots & \vdots \\
    0 & 0 & \hdots & \pm 1
 \end{pmatrix}
 \label{eq:spread_mat}
\end{equation}
As stated in \cite{TR_FD_TD}, the idea behind the spreading is that up-sampling a signal in the TD is equivalent to the repetition and shifting of its spectrum in the FD. In doing so, each data symbol will be transmitted onto $U$ different subcarriers with a spacing of $N$ subcarriers, introducing frequency diversity. The spread sequence is then precoded before being transmitted. This requires the knowledge of Bob channel frequency response (CFR) at Alice.  The channels between Alice and Bob ($\uu{H}_{\text{B}}$) and between Alice and Eve ($\uu{H}_{\text{E}}$) are assumed to be static during the transmission of one OFDM symbol. $\uu{H}_{\text{B}}$ and $\uu{H}_{\text{E}}$ are $Q\times Q$ diagonal matrices whose elements are $H_{\text{B},q}$ and $H_{\text{E},q}$ (for $q = 0,...,Q-1$) and follow a zero-mean unit-variance normal distribution. The precoding matrix $\uu{H}_{\text{B}}^*$ is also a diagonal matrix with elements $H_{\text{B},q}^*$. At the receiver, a despreading operation is performed by applying $\uu{S}^H$. We consider that Bob and Eve decoding abilities are identical. They both know the spreading sequence and apply a Zero Forcing (ZF) equalization. A perfect synchronization is assumed at Bob and Eve positions.

\subsubsection{Received sequence at the intended position}
After despreading, the received sequence at Bob is:
\begin{equation}
    \underline{Y}_{\text{B}} = \underline{\underline{S}}^H  \left|\underline{\underline{H}}_{\text{B}} \right|^2\underline{\underline{S}}\; \underline{X} +  \underline{\underline{S}}^H \underline{V}_\text{B} 
    \label{eq:rx_bob}
\end{equation}
where $\underline{V}_\text{B}$ is the FD complex additive white Gaussian noise (AWGN). The noise's auto-correlation is $\mathbb{E}[ \left|V_{\text{B},n}\right|^2 ]  = \sigma_{\text{V,B}}^2$ and the covariance matrix is $\mathbb{E}[(\uu{S}^H  \underline{V}_\text{B}) . (\uu{S}^H \underline{V}_\text{B})^H ] = \sigma_{\text{V,B}}^2 . \uu{I}_N$. We also assume that the signal $X_n$ and noise $V_{\text{B,n}}$ are independent of each other. In (\ref{eq:rx_bob}), each transmitted symbol is affected by a real gain at the position of the legitimate receiver since the product $\uu{H}_{\text{B}}\;\uu{H}^*_{\text{B}}$ is a real diagonal matrix. The gains differ between each symbol in the OFDM block but increases with an increase of the BOR value as each symbol would be sent on more subcarriers and would benefit from a larger frequency diversity gain. If we consider a fixed bandwidth, the TR focusing effect is enhanced for higher BOR's at the expense of the data rate. After ZF equalization, we obtain:
\begin{equation}
\resizebox{0.91\hsize}{!}{$
    \underline{\hat{X}}_{\text{B}} = \left(\underline{\underline{S}}^H  \left|\underline{\underline{H}}_{\text{B}} \right|^2\underline{\underline{S}}\right)^{-1} \left(\underline{\underline{S}}^H  \left|\underline{\underline{H}}_{\text{B}} \right|^2\underline{\underline{S}}\; \underline{X} +  \underline{\underline{S}}^H \underline{V}_\text{B}\right)$}
    \label{eq:rx_bob_eq}
\end{equation}
From (\ref{eq:rx_bob_eq}), we observe that the transmit data is perfectly recovered in the absence of noise.

\subsubsection{Received sequence at the unintended position}
After despreading, the data received at the unintended position is given by:
\begin{equation}
    \underline{Y}_{\text{E}}= \underline{\underline{S}}^H  \underline{\underline{H}}_{\text{E}} \underline{\underline{H}}^*_{\text{B}}  \underline{\underline{S}}\; \underline{X}  +  \underline{\underline{S}}^H  \underline{V}_\text{E}
    \label{eq:rx_eve}
\end{equation}
where $\underline{V}_\text{E}$ is the complex AWGN. The noise's auto-correlation is $\mathbb{E}[ \left|V_{\text{E},n}\right|^2 ] = \sigma_{\text{V,E}}^2$ and the covariance matrix is $\mathbb{E}[(\uu{S}^H  \underline{V}_\text{E}) . (\uu{S}^H \underline{V}_\text{E})^H ] = \sigma_{\text{V,E}}^2 . \uu{I}_N$.  In (\ref{eq:rx_eve}), $\underline{\underline{H}}_{\text{E}} \underline{\underline{H}}^*_{\text{B}}$ is a complex diagonal matrix, and each transmitted symbol is affected by a random complex coefficient. The magnitude of this coefficient does not depend on the BOR value. It results in an absence of TR gain at the unintended position. As a consequence, worse decoding performance is obtained compared to the intended position. Eve needs lower noise power than Bob to reach the same bit-error-rate (BER). After ZF equalization one obtains:
\begin{equation}\resizebox{0.91\hsize}{!}{$
    \underline{\hat{X}}_{\text{E}} = \left(\underline{\underline{S}}^H  \underline{\underline{H}}_{\text{E}} \underline{\underline{H}}^*_{\text{B}}\underline{\underline{S}}\right)^{-1} \left(\underline{\underline{S}}^H  \underline{\underline{H}}_{\text{E}} \underline{\underline{H}}^*_{\text{B}}\underline{\underline{S}}\; \underline{X} +  \underline{\underline{S}}^H \underline{V}_\text{E}\right)$}
    \label{eq:rx_eve_eq}
\end{equation}
Equation (\ref{eq:rx_eve_eq}) shows that in the classical FD TR SISO OFDM communication scheme, the data could potentially be recovered at Eve's position. A similar BER could be obtained at Eve if she is close to Alice than Bob is and/or if the noise power is less than Bob's one. This motivates the addition of AN in order to corrupt the data detection at any unintended positions and to secure the communication.

\subsection{FD TR SISO OFDM communication with Artificial Noise}

\begin{figure}[t]
    \centering
    \centerline{\includegraphics[width = .53\textwidth]{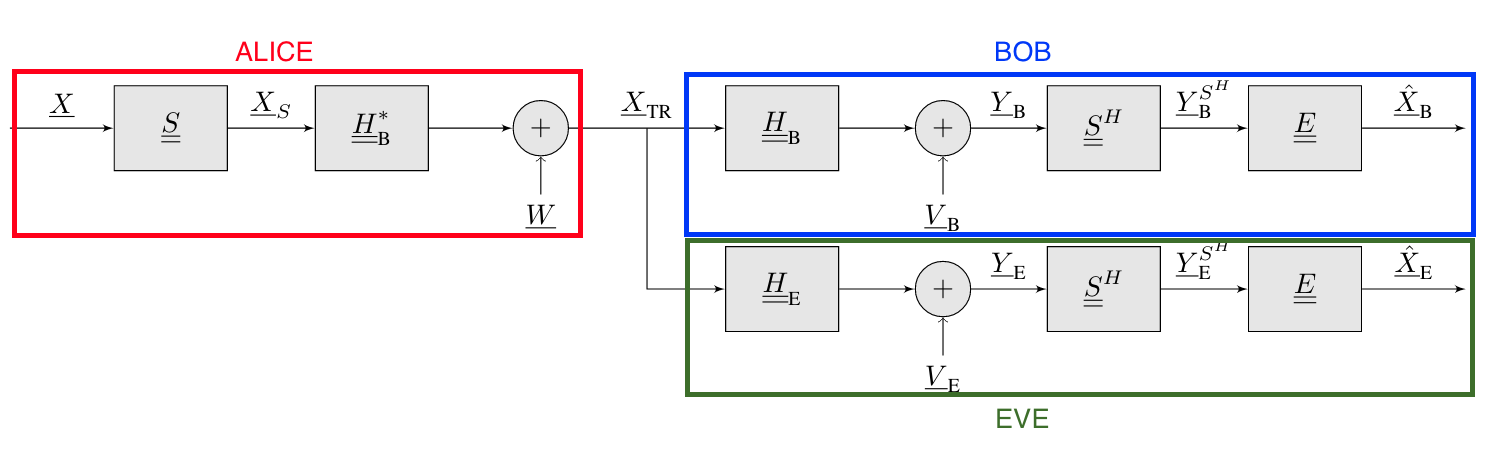}}
    \caption{FD TR SISO OFDM system with added Artificial Noise}
    \label{fig:TR_FD_AN}
\end{figure}
In order to further secure the communication between Alice and Bob, an AN signal $\underline{W}$ is added after precoding to the useful signal $\underline{X}_s$ at the transmitter side, as depicted in Fig.\ref{fig:TR_FD_AN}. The AN should not have any impact at Bob's position but should be seen as interference everywhere else since Alice does not have any information about Eve's CSI. Furthermore, this signal should not be guessed at the unintended positions to ensure the secure communication. With these considerations, the transmitted sequence becomes:
\begin{equation}
    \underline{X}_{\text{TR}} = \sqrt{\alpha} \;\underline{\underline{H}}_{\text{B}}^*  \underline{\underline{S}}\; \underline{X} +  \sqrt{1-\alpha} \; \underline{W}
    \label{eq:sym_rad_AN}
\end{equation} 
where $\alpha \in [0,1]$ defines the ratio of the total power dedicated to the useful signal, knowing that $\mathbb{E}\left[ \left|\underline{\underline{H}}_{\text{B}}^*  \underline{\underline{S}}\; \underline{X}  \right|^2 \right] = \mathbb{E}\left[ \left|\underline{W} \right|^2 \right]$. Whatever the value of $\alpha$, the total transmitted power remains constant.

\subsubsection{AN design}
In order not to have any impact at the intended position, the AN signal must satisfy the following condition:
\begin{equation}
    \underline{\underline{S}}^H\underline{\underline{H}}_{\text{B}} \underline{W} \; = \; \underline{0}
    \label{eq:an_cond}
\end{equation}
where $\underline{0}$ is the null vector of dimension $N\times 1$. From (\ref{eq:an_cond}), the following system must be solved:
\begin{equation}
\resizebox{1.3\hsize}{!}{
    $
    \underset{\begin{array}{c}\\ N\times Q \end{array}}%
{%
    \begin{pmatrix}
    \pm 1 & 0 & \hdots & 0 & & \pm 1 & 0 & \hdots & 0  \\
    0 & \pm 1 & \hdots & 0 & \hdots & 0 & \pm 1 & \hdots & 0 \\
    \vdots & & \ddots & \vdots & \hdots & \vdots & & \ddots & \vdots \\
    0 & 0 & \hdots & \pm 1  & &0 & 0 & \hdots & \pm 1\\
    \end{pmatrix}
}
\underset{\begin{array}{c}\\ Q\times 1 \end{array}}%
{%
    \begin{pmatrix}
     H_{\text{B},0} \\
    H_{\text{B},1} \\
    \vdots \\
    H_{\text{B},Q-1}
    \end{pmatrix}
}
    \odot
\underset{\begin{array}{c}\\ Q\times 1 \end{array}}%
{%
    \begin{pmatrix}
    W_0 \\
    W_1 \\
    \vdots\\
    W_{Q-1}
    \end{pmatrix}
}
    = 
\underset{\begin{array}{c}\\ N\times 1 \end{array}}%
{%
    \begin{pmatrix}
    0\\
    0\\
    \vdots\\
    0
    \end{pmatrix}
}$}
    \label{eq:AN_generation}
\end{equation}
where ``$\odot$" represents the element-wise (Hadamard) product. Equation (\ref{eq:AN_generation}) is a set of $N$ equations with $Q$ unknowns. Since $Q = NU$, as soon as $U\geq 2$, (\ref{eq:AN_generation}) becomes under-determined and the AN vector can be generated from a set of infinite possibilities. Let us define $\uu{S}^H = [\uu{S}_0 \; \uu{S}_1 \; ... \; \uu{S}_{U-1}]$, where $\uu{S}_i$ is the $i^{\text{th}}$ diagonal matrix of $\uu{S}^H$ of size $N\times N$. We define as $S_{i,p}$ as the $p^{th}$ diagonal element of $\uu{S}_i$. If we denote the $i^{\text{th}}$ diagonal element of $\uu{H}_{\text{B}}$ as $H_i$, (\ref{eq:AN_generation}) becomes: 
\begin{equation}
\setlength{\jot}{15pt}
\begin{split}
\left\{
\begin{array}{lll}
    \sum\limits_{i=0}^{\small{U-1}} S_{i,0} H_{iN} W_{iN} &=& 0 \\[2pt]
    \sum\limits_{i=0}^{U-1} S_{i,1}H_{iN+1} W_{iN+1} &=& 0 \\[2pt]
    \hspace{.8in}\vdots  && \vdots \\
    \sum\limits_{i=0}^{U-1} S_{i,N-1}H_{iN +N-1} W_{iN+N-1} &=& 0
\end{array}
\right.
\end{split}
\label{eq:system_AN_to_determine}
\end{equation}
In (\ref{eq:system_AN_to_determine}), each equation consists in a sum of $U$ elements, with the channel components and the spreading sequence being known. The idea is to generate $U-1$ components of the AN vector in every equations of (\ref{eq:system_AN_to_determine}) as normal random variables, and to determine the last AN components that ensures (\ref{eq:an_cond}) as follow:
\begin{equation}
\setlength{\jot}{5pt}
\begin{split}
\left\{
\begin{array}{lll}
    W_{Q-N}  &=& - \frac{\sum\limits_{i=0}^{U-2}  S_{i,0} H_{iN} W_{iN}}{ S_{U-1,0} H_{Q-N}} \\[5pt]
    W_{Q-N+1}& =&  - \frac {\sum\limits_{i=0}^{U-2}  S_{i,1}H_{iN+1} W_{iN+1}}{ S_{U-1,1}H_{Q-N+1}} \\[2pt]
    \hspace{.2in} \vdots  &&  \hspace{.4in} \vdots \\[2pt]
    W_{Q-1} &=&  - \frac{\sum\limits_{i=0}^{U-2}  S_{i,N-1}H_{iN +N-1} W_{iN+N-1}}{ S_{U-1,N-1}H_{Q-1}}
\end{array}
\right.
\end{split}
\label{eq:system_AN_determined}
\end{equation}

\subsubsection{Received sequence at the intended position}
After despreading, the received sequence at Bob is: 
\begin{equation}
    \underline{Y}_{\text{B}} = \sqrt{\alpha} \; \underline{\underline{S}}^H \left|\underline{\underline{H}}_{\text{B}} \right|^2 \underline{\underline{S}}\; \underline{X} \;  +  \;  \underline{\underline{S}}^H \underline{V}_\text{B} 
    \label{eq:rx_bob_AN}
\end{equation}
Again, each transmitted symbol is affected by a real gain depending on the BOR value and weighted by $\sqrt{\alpha}$. One can observe that no AN contribution is present in (\ref{eq:rx_bob_AN}) since (\ref{eq:an_cond}) is respected. A ZF equalization is performed at the receiver:
\begin{equation}
 \resizebox{0.89\hsize}{!}{
    $\hat{\underline{X}}_{\text{B}} = \left( \sqrt{\alpha} \; \underline{\underline{S}}^H \left|\underline{\underline{H}}_{\text{B}} \right|^2 \underline{\underline{S}} \right)^{-1}  \left(\sqrt{\alpha} \; \underline{\underline{S}}^H \left|\underline{\underline{H}}_{\text{B}} \right|^2 \underline{\underline{S}}\; \underline{X} \;  +  \;  \underline{\underline{S}}^H \underline{V}_\text{B}\right)$} 
    \label{eq:rx_bob_AN_eq}
\end{equation}
From (\ref{eq:rx_bob_AN_eq}), a perfect data recovery is possible in the absence of noise.

\subsubsection{Received sequence at the unintended position}
After despreading, the received sequence at the unintended position is given by:
\begin{equation}
\resizebox{0.89\hsize}{!}{$
    \underline{Y}_{\text{E}} = \sqrt{\alpha} \; \underline{\underline{S}}^H \underline{\underline{H}}_{\text{E}} \underline{\underline{H}}^*_{\text{B}} \underline{\underline{S}}\; \underline{X} \; +  \; \sqrt{1-\alpha} \; \underline{\underline{S}}^H \underline{\underline{H}}_{\text{E}} \underline{W}  \; +  \; \underline{\underline{S}}^H  \underline{V}_\text{E} $}
    \label{eq:rx_eve_an}
\end{equation}
In (\ref{eq:rx_eve_an}), a term depending on the AN signal appears since $ \underline{\underline{S}}^H\underline{\underline{H}}_{\text{E}} \underline{W} \neq \underline{0}$. This term introduces an interference at Eve and thus scrambles the received constellation even in a noiseless environment. After ZF equalization, the estimated symbols are:
\begin{equation}
\resizebox{0.89\hsize}{!}{$
    \begin{split}
         \hat{\underline{X}}_{\text{E}} &= \left( \sqrt{\alpha} \; \underline{\underline{S}}^H \underline{\underline{H}}_{\text{E}} \underline{\underline{H}}^*_{\text{B}} \underline{\underline{S}} \right)^{-1} \\
         & \left( \sqrt{\alpha} \;\underline{\underline{S}}^H \underline{\underline{H}}_{\text{E}} \underline{\underline{H}}^*_{\text{B}} \underline{\underline{S}}\; \underline{X} \; +  \; \sqrt{1-\alpha} \; \underline{\underline{S}}^H \underline{\underline{H}}_{\text{E}} \underline{W}  \; +  \; \underline{\underline{S}}^H  \underline{V}_\text{E}  \right)
    \end{split}$}
    \label{eq:rx_an_eve_eq}
\end{equation}
Equation (\ref{eq:rx_an_eve_eq}) shows that the addition of AN in the FD TR SISO OFDM communication can secure the data transmission. It is to be noted that since $\underline{W}$ is generated from an infinite set of possibilities, even if Eve happens to know $\uu{H}_{\text{B}}$, she cannot estimate the AN to try retrieving the data.  The degree of security will depend on the amount of AN energy that is injected into the communication, as shown in Section \ref{sec:perf}.


\section{Performance Assessment}
\label{sec:perf}
The secrecy rate (SR) is defined as the maximum transmission rate that can be supported by the legitimate receiver's channel while ensuring the impossibility for the eavesdropper to retrieve the data, \cite{TR_Tran_secrecy_capa}. In the ergodic sense, it can be expressed as:
\begin{equation}
\begin{split}
    C_S &=  \mathop{\mathbb{E}} \left[\log_2{\left(1+\gamma_B\right)} - \log_2{\left(1+\gamma_E\right)}\right] \; \; \; , \; \; \;  \gamma_B > \gamma_E \\
    &\leq   \log_2 \left( 1+ \mathop{\mathbb{E}}\left[\gamma_B\right] \right) - \log_2 \left( 1+ \mathop{\mathbb{E}}\left[\gamma_E \right] \right) 
    \end{split}
    \label{eq:SR}
\end{equation}
with $\gamma_B$ and $\gamma_E$ being respectively the signal-to-interference-plus-noise Ratio (SINR) at Bob and Eve's positions. The inequality in (\ref{eq:SR}) arises from the Jensen's inequality.

\subsection{SINR determination}

\subsubsection{At the intended position}
At Bob, the received signal after despreading is given by (\ref{eq:rx_bob_AN}). Using the Jensen's inequality, a lower bound on the average SINR can be derived for the transmitted symbols $n$ as:
\begin{equation}
\begin{split}
    \mathop{\mathbb{E}} \left[\gamma_{B,n}\right] &= \mathop{\mathbb{E}} \left[ \frac{  \alpha \left| K_n \; X_n\right|^2  }{  \left| V_{B,n} \right|^2}  \right]  = \alpha \mathop{\mathbb{E}} \left[ \left| K_n \; X_n\right|^2 \right]  \mathop{\mathbb{E}} \left[ \frac{1}{\left| V_{B,n} \right|^2} \right]  \\
    & \geq  \frac{\alpha \mathop{\mathbb{E}} \left[ \left| K_n \; X_n\right|^2 \right] }{\mathop{\mathbb{E}} \left[ \left| V_{B,n} \right|^2 \right]} =  \frac{\alpha \mathop{\mathbb{E}} \left[ \left| K_n \; X_n\right|^2 \right] }{\sigma^2_{\text{V,B}}}
    \label{eq:RV_sinr_b}
\end{split}
\end{equation}
where $K_n = \frac{1}{U}\; \sum_{i=0}^{U-1} \left| H_{\text{B}, n + iN}\right|^2$ is a real random variable (RV) independent of the data symbol $X_n$, and the noise $V_{B,n}$ is considered independent of $K_n$ and $X_n$. If $H_{\text{B}, n + iN}$ ($\forall i = 0,...,U-2$) elements are assumed to be non-correlated\footnote{Thanks to the design of the spreading matrix, the $U$ subcarriers composing one symbol are spaced by $N = Q/U$ subcarriers. If this distance is larger than the coherence bandwidth of the channel, the assumption holds. This usually occurs in rich multipath environments and for sufficiently large bandwidths and moderate BOR values.}, $K_n$   can be approximated as following a chi-square  distribution with $U$ degrees of freedom, so that: 
\begin{equation}
    \mathop{\mathbb{E}}\left[ \left| K_n \right|^2\right] = \int \limits_{0}^{\infty} z^2  f_Z(z) dz  = \frac{(U+1)}{U}
\end{equation}
 Furthermore, remembering that $\mathop{\mathbb{E}}\left[ \left| X_n \right|^2\right] = 1$, the average SINR for symbols $n$ at the intended position is then given by:
\begin{equation}
    \mathop{\mathbb{E}} \left[\gamma_{B,n}\right] \geq \frac{\alpha \;(U+1)}{U \; \sigma_{\text{V,B}}^2}
    \label{sinr_bob}
\end{equation}
It was observed in simulations than the upper-bound (\ref{sinr_bob}) is tight enough to be used as an approximation of the averaged SINR at the intended position.

\subsubsection{At the unintended position}
\label{subsub:unintended_sinr}
The received signal after despreading is (\ref{eq:rx_eve_an}). Let's introduce $A_1  = \sqrt{1-\alpha} \underline{\underline{S}}^H \underline{\underline{H}}_{\text{E}} \underline{W}$ and $A_2 = \sqrt{\alpha}  \underline{\underline{S}}^H  \underline{\underline{H}}_{\text{E}}  \underline{\underline{H}}^*_{\text{B}} \underline{\underline{S}}\; \underline{X}$. Using the Jensen's inequality, an approximation of a lower bound of the averaged SINR of the symbols $n$ at the unintended position can be derived as\footnote{Neglecting the covariance between $\left|A_{2,n}\right|^2$ and $\left| V_{E,n} + A_{1,n}\right|^2$, as  done in the first line of (\ref{eq:expected_sinr_eve}), makes the nature of the bound, i.e., lower or upper, obtained for $\mathop{\mathbb{E}} \left[\gamma_{E,n}\right]$ uncertain. However, we have observed by simulations that it remains a lower one for all considered scenarios.}:
\begin{equation}
\begin{split}
    \mathop{\mathbb{E}} \left[\gamma_{E,n}\right] &= \mathop{\mathbb{E}} \left[ \frac{  \left| A_{2,n} \right|^2  }{  \left| V_{E,n} + A_{1,n} \right|^2 }  \right]  \approx  \mathop{\mathbb{E}} \left[ \left| A_{2,n} \right|^2 \right]  \mathop{\mathbb{E}} \left[ \frac{1}{ \left| V_{E,n} + A_{1,n} \right|^2} \right]  \\
    & \geq \frac{\mathop{\mathbb{E}} \left[ \left| A_{2,n} \right|^2 \right] }{\mathop{\mathbb{E}} \left[\left| V_{E,n} + A_{1,n} \right|^2 \right]} =  \frac{\mathop{\mathbb{E}} \left[ \left| A_{2,n} \right|^2 \right] }{\mathop{\mathbb{E}} \left[\left| V_{E,n} \right|^2 \right] +  \mathop{\mathbb{E}} \left[\left|A_{1,n} \right|^2 \right]}
    \label{eq:expected_sinr_eve}
\end{split}
\end{equation}
Assuming $H_{\text{E}, n + iN}$ ($\forall i = 0,...,U-2$) elements non-correlated and neglecting the correlation introduced in $\underline{W}$ by (\ref{eq:system_AN_determined})\footnote{This approximation holds for sufficiently large BOR values since only one term depends on $U-1$ terms in each equation of (\ref{eq:system_AN_determined}).}, the AN interference can be calculated as:
\begin{equation}
\resizebox{0.89\hsize}{!}{$
    \begin{split}
        \mathop{\mathbb{E}}\left[ \left| A_{1,n} \right|^2 \right]
        &= \frac{(1-\alpha)}{U} \; \sum_{i=0}^{U-1} \mathop{\mathbb{E}}\left[\left| W_{n+iN} \right|^2 \right]  \; \underbrace{\mathop{\mathbb{E}}\left[\left|  H_{\text{E}, n + iN} \right|^2 \right]}_{=1} \\
        &= (1-\alpha)\; \sigma^2_{\text{AN}}
    \end{split}$}
\end{equation}
where $\sigma^2_{\text{AN}} = \mathbb{E}\left[\left| W_{n+iN} \right|^2 \right]$, and $W_{n+iN}$ and $H_{\text{E}, n + iN}$ are independent.\\
The energy related to the useful symbol is:
\begin{equation}
\resizebox{0.89\hsize}{!}{$
    \begin{split}
         \mathop{\mathbb{E}}\left[ \left| A_{2,n} \right|^2 \right]
         &= \frac{\alpha}{U^2}\; \mathop{\mathbb{E}}\left[ \left| \sum_{i=0}^{U-1}  H_{\text{E},n+iN} \; H^*_{\text{B},n+iN} \right|^2 \right] \; \underbrace{\mathop{\mathbb{E}}\left[ \left| X_{n+iN}\right|^2 \right]}_{=1} \\
         &= \frac{\alpha}{U^2} \; \mathop{\mathbb{E}}\left[ \left| Z_n \right|^2 \right]
    \end{split}$}
    \label{eq:expected_value_eve}
\end{equation}
with $Z_n = \sum_{i=0}^{U-1} Z_{n,i}$ and $Z_{n,i} = H_{\text{E},n+iN}.H_{\text{B},n+iN}^*$ (where $H_{\text{E},n+iN}$ and $H_{\text{B},n+iN}^*$ are assumed to be statistically independent and identically distributed (i.i.d.) complex Gaussian RVs). $Z_n$, similarly to $K_n$, is  the sum of uncorrelated complex RVs. Introducing $R = | Z_n |$, we obtain the PDF of $R$ as in \cite{TR_FD_TD}:
\begin{equation}
    f_R(r) = \frac{4r^U}{\Gamma(U)} \; \mathbb{K}_{U-1}\left( 2r\right)
    \label{eq:pdf_eve}
\end{equation}
where $\Gamma(U) = (U-1)! = \int_0^\infty z^{U-1} e^{-z} dz$ is the Gamma function of the integer $U$. In (\ref{eq:pdf_eve}), $\mathbb{K}_{U}$ is the $U^{\text{th}}$ order modified Bessel function of the second kind which can be approximated by:
\begin{equation}
    \mathbb{K}_{U}(x) \approx \sum_{q=0}^{D}\sum_{l=0}^{D} \psi(U,l,q) \; e^{-x} \; x^{q-U}
\end{equation}
where $D$ specifies the number of expansion terms and $\psi(U,l,q)$ is given by:
\begin{equation}
    \psi(U,l,q) = \frac{(-1)^q \; \sqrt{\pi} \; \Gamma(2U) \; \Gamma(1/2+l-U) \; \mathbb{L}(l,q)}{2^{U-q}\; \Gamma(1/2-U) \; \Gamma(1/2+l+U) \; l!}
\end{equation}
where $\mathbb{L}(l,q)$ is the Lah number \cite{TR_la} with the conventions $\mathbb{L}(0,0) = 1$,  $\mathbb{L}(l,0) = 0$,  $\mathbb{L}(l,1) = l! \; \forall l > 0$. Eq. (\ref{eq:expected_value_eve}) therefore becomes: 
\begin{equation}
\resizebox{0.89\hsize}{!}{$
    \begin{split}
        \mathop{\mathbb{E}}\left[ \left| A_{2,n} \right|^2 \right] &= \frac{\alpha}{U^2} \int\limits_{0}^{\infty} r^2 \;\frac{ 4r^U}{\Gamma(U)}\; \mathbb{K}_{U-1}(2r) \; dr \\
        &= \frac{4\alpha}{U^2 \; (U-1)!} \; \sum_{q=0}^{D}\sum_{l=0}^{D} \psi(U-1,l,q) \int\limits_{0}^{\infty} r^{U+2} \; e^{-2r}  \;  (2r)^{q-U+1} \; dr \\
        &=  \frac{4\alpha}{U^2 \; (U-1)!\; 2^{U+3}} \sum_{q=0}^{D}\sum_{l=0}^{D} \psi(U-1,l,q) \; \Gamma(q+4) 
    \end{split}$}
\end{equation}
Consequently, an approximation of the averaged SINR's lower bound for transmitted symbols $n$ at the unintended position can be expressed as:
\begin{equation}
    \mathop{\mathbb{E}} \left[\gamma_{E,n}\right] \gtrapprox \frac{\frac{4\alpha}{U^2  (U-1)! 2^{U+3}} \sum_{q=0}^{D}\sum_{l=0}^{D} \psi(U-1,l,q)  \Gamma(q+4)}{\sigma^2_{\text{V,E}} \; + \; (1-\alpha) \sigma^2_{\text{AN}} }
    \label{sinr_eve}
\end{equation}
It has been observed in simulations that (\ref{sinr_eve}) remains a lower bound for $\gamma_{E,n}$ as long as $U\geq4$. Furthermore, it is a tight-enough bound to be used as an approximation for the secrecy capacity derivation.

\subsection{Optimal amount of Artificial Noise energy to maximize the secrecy capacity}
With (\ref{eq:SR}), (\ref{sinr_bob}) and (\ref{sinr_eve}), it is possible to obtain a closed-form approximation of the SR upper bound and determine the amount of AN energy to inject that maximizes the SR.

Introducing $A = \frac{1}{U^2 \; (U-1)!\; 2^{U+3}} \sum_{q=0}^{D}\sum_{l=0}^{D} \psi(U-1,l,q) \; \Gamma(q+4)$, the SR is therefore:
\begin{equation}
\resizebox{0.89\hsize}{!}{$
\begin{split}
    C_s &\lessapprox \log_2 \left(1 + \frac{\alpha (U+1)}{U \sigma_{\text{V,B}}^2} \right)  - \log_2\left( 1 + \frac{4\alpha A }{\sigma^2_{\text{V,}} \; + \; (1-\alpha) \sigma^2_{\text{AN}} } \right)\\
    &\lessapprox \log_2 \left( \frac{U \sigma_{\text{V,B}}^2 \; + \; \alpha (U+1)}{U  \sigma_{\text{V,B}}^2} \; . \; \frac{\sigma^2_{\text{V,E}} \; + \; (1-\alpha) \sigma^2_{\text{AN}}}{\sigma^2_{\text{V,E}} \; + \; (1-\alpha) \sigma^2_{\text{AN}} \; + \; 4\alpha A } \right)
    \label{eq:SR_anal1}
\end{split}$}
\end{equation}
Let us denote $T_1 = \sigma^2_{\text{AN}}(U+1)$, $T_2 = \sigma^2_{\text{V,E}}(U+1) -  U \sigma_{\text{V,B}}^2\sigma^2_{\text{AN}}  + \sigma^2_{\text{AN}}(U+1)$, $T_3 = U \sigma_{\text{V,B}}^2 \left[ \sigma^2_{\text{V,E}}  +  \sigma^2_{\text{AN}}\right]$ and $T_4 = 4A U \sigma_{\text{V,B}}^2 -  \sigma_{\text{V,B}}^2\sigma^2_{\text{AN}}$. After some manipulations, (\ref{eq:SR_anal1}) becomes:
\begin{equation}
    C_s \lessapprox \log_2 \left( \frac{-\alpha^2 T_1 \; + \; \alpha T_2 \; + \; T_3}{\alpha T_4 \; + \; T_3} \right)
    \label{eq:SR_anal2}
\end{equation}
To  maximize the secrecy rate as a function of the parameter $\alpha$, we find the zeroes of:
\begin{equation}
\begin{split}
    \frac{\partial C_s}{\partial \alpha} &= \frac{ \frac{-\alpha^2 T_1 T_4 \; - \; 2 \alpha T_1 T_3 \; + \; \left( T_2 T_3 \; - \; T_3 T_4 \right) }{\left( \alpha T_4 \; + \; T_3\right)^2} }{ \frac{-\alpha^2 T_1 \; + \; \alpha T_2 \; + \; T_3}{\alpha T_4 \; + \; T_3} \; . \; \ln{2}} 
    \label{eq:SR_derivative}
\end{split}
\end{equation}
After some algebraic manipulations, one obtains: 
\begin{equation}
\resizebox{0.89\hsize}{!}{$
         \frac{\partial C_s}{\partial \alpha} = 0
         \; \Leftrightarrow \; \alpha_{opt} = \frac{\pm\sqrt{T_1^2 T_3^2 \; + \; T_1 T_2 T_3 T_4 \; - \; T_1 T_3 T_4^2} \; - \; T_1 T_3}{T_1 T_4}$}
         \label{eq:best_alpha}
\end{equation}
where only the positive roots are solutions since $\alpha \in [0,1]$.

\section{Simulation Results}
\begin{figure}[t]
    \centering
    \centerline{\includegraphics[width = .4\textwidth]{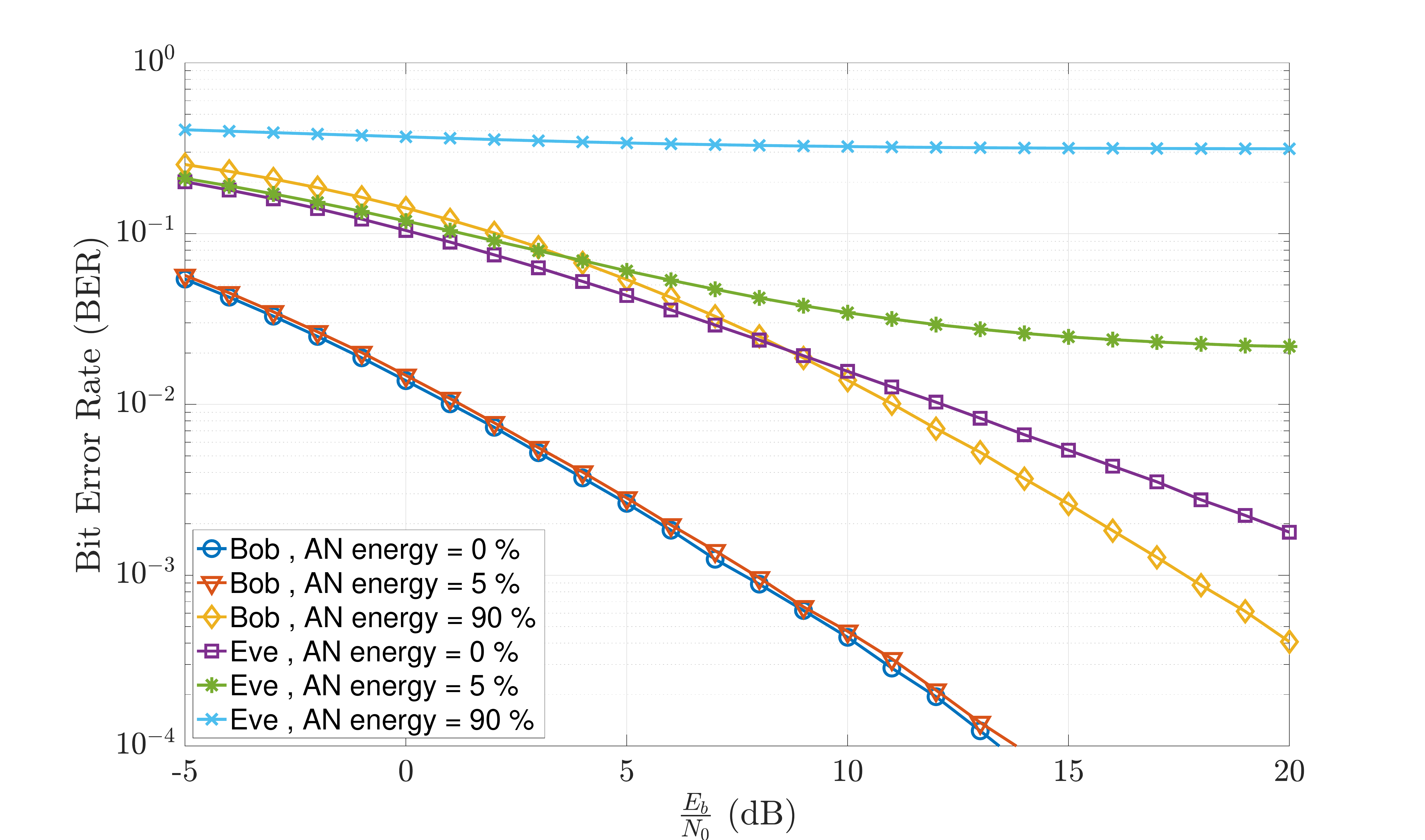}}
    \caption{BER as a function of the level of noise for different AN energy values, BOR = 4}
    \label{fig:ber_ebno}
\end{figure}

\begin{figure}[t]
    \centering
    \centerline{\includegraphics[width = .4\textwidth]{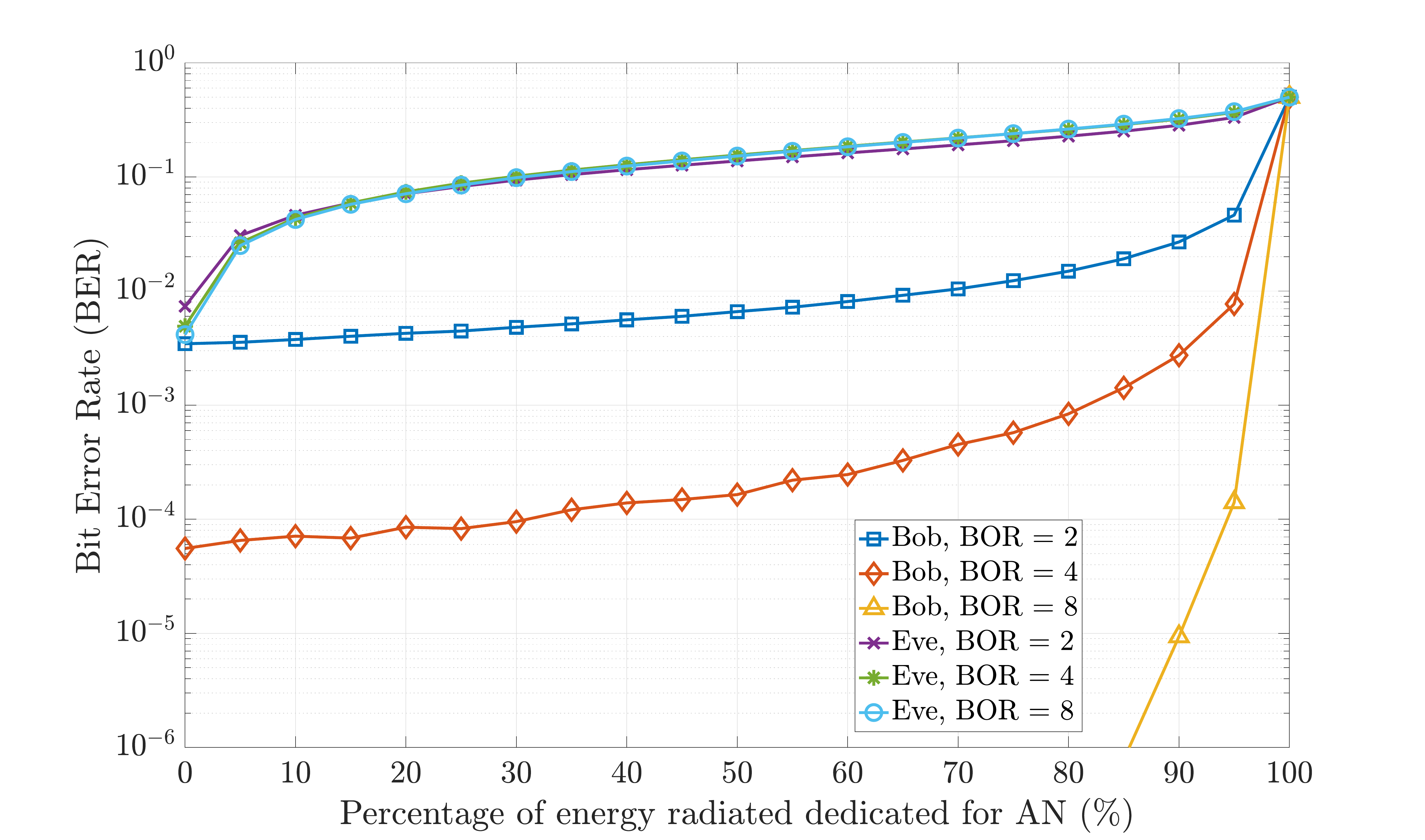}}
    \caption{BER as a function of AN energy for different BOR values, $E_b/N_0 = 15$dB}
    \label{fig:ber_alpha}
\end{figure}

\begin{figure}[t]
    \centering
    \centerline{\includegraphics[width = .43\textwidth]{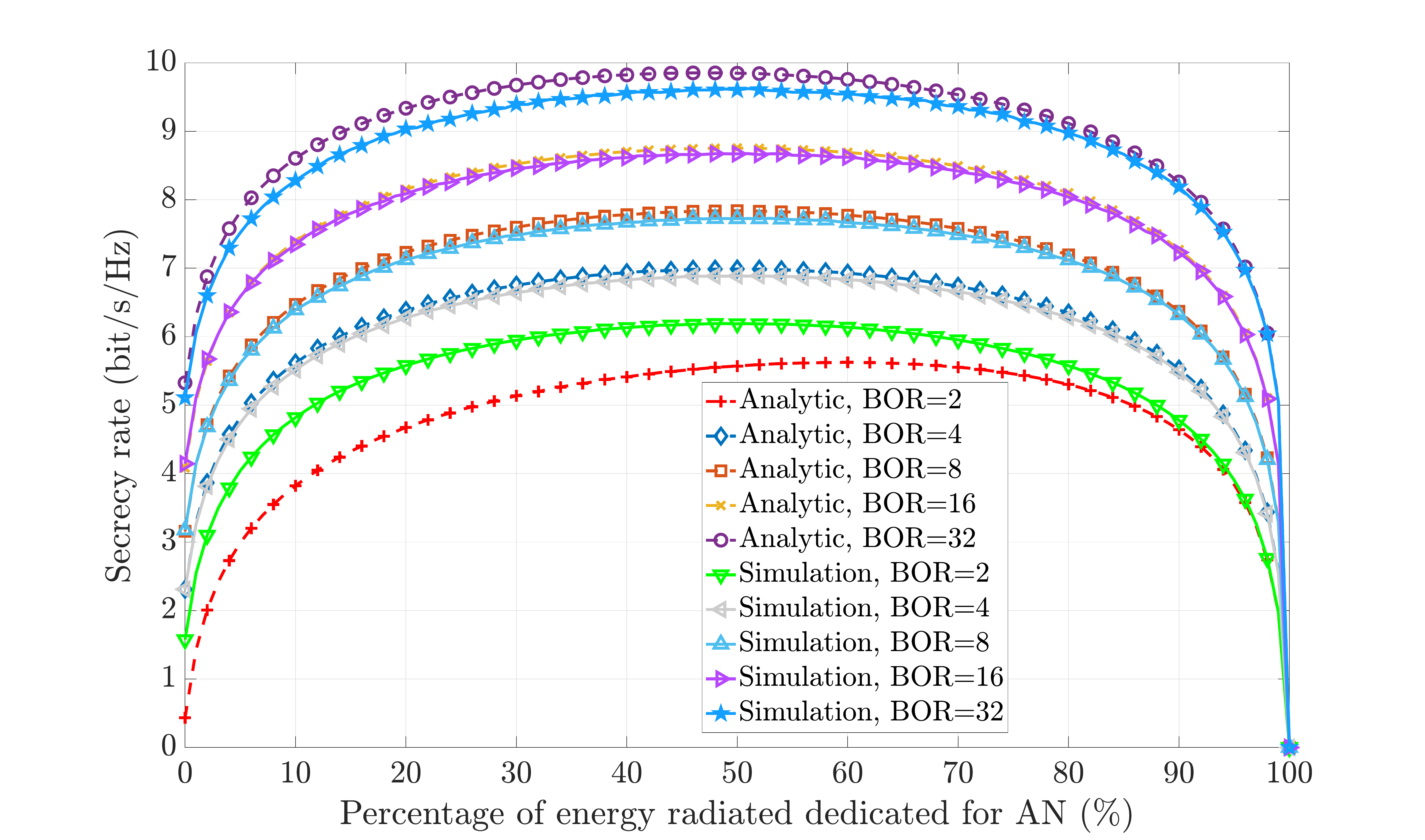}}
    \caption{Secrecy Rate as a function of the AN energy for different BOR values, analytic vs simulated curves $E_b/N_0=20$ dB }
    \label{fig:secrecy_alpha_bor}
\end{figure}

\label{sec:result}
A 256-subcarrier SISO OFDM system is considered. Bob and Eve channels are assumed to be uncorrelated. Each subcarrier is Rayleigh distributed and there is no correlation between subcarriers. The overall channel energies are normalized to unity for each channel realization. Bob's CSI is assumed to be perfectly known at Alice. Bob and Eve have the same level of noise. The number of expansion terms of the modified Bessel function is set to $D=20$ for which convergence has been observed. Simulations with 100 channel realizations and 300 OFDM blocks were performed using a 4-QAM modulation scheme. 

\subsection{Decoding results}
Fig. \ref{fig:ber_ebno} and \ref{fig:ber_alpha} show the system performance in terms of BER obtained after ZF equalization at Bob and Eve. In Fig. \ref{fig:ber_ebno}, the BER is plotted as a function of $E_b/N_0$, where $E_b$ is the energy per bit, calculated after spreading, and $N_0$ is the noise power spectral density.  Different levels of AN energy are investigated at fixed BOR. It can be observed that, as soon as a small amount of radiated energy is dedicated to AN, e.g., $5\%$, Eve's BER strongly increases. At the intended position, the BER also increases but much slower. The reason is that the higher the percentage of energy dedicated to AN, the lower the received useful signal power at Bob. In Fig. \ref{fig:ber_alpha}, the BER is plotted as a function of the AN energy, at fixed $E_b/N_0=15$ dB and different BOR values. At the unintended position, the BER naturally increases with the amount of injected AN, whatever the BOR value. At Bob, low BER values can be maintained for high AN power by increasing the BOR, as anticipated from Section \ref{subsec:traditional_FDTR}. One can notice that, when $\alpha \to 0$, the BER curves all converge to $0.5$, as expected.

\subsection{Secrecy results}

\begin{figure}[t]
    \centering
    \centerline{\includegraphics[width = .37\textwidth]{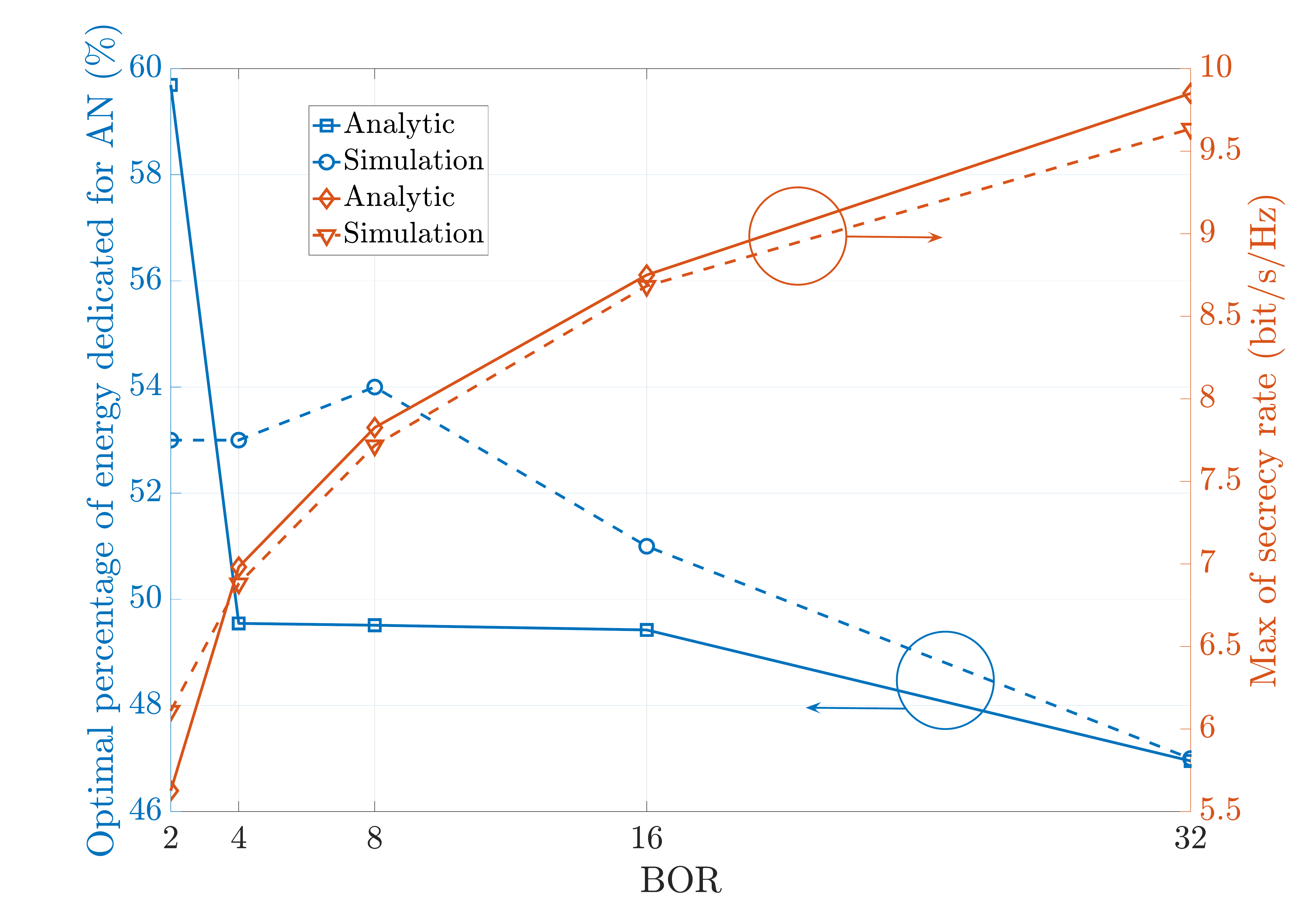}}
    \caption{Optimal AN energy to inject and maximal Secrecy Rate for different BOR values, $E_b/N_0=20$ dB}
    \label{fig:secrecy_alpha_bor_optimal}
\end{figure}

Fig. \ref{fig:secrecy_alpha_bor} shows the SR evolution as a function of $\alpha$ for different BOR values. First, it can be seen that analytic curves, given by (\ref{eq:SR_anal2}), approximate well the simulation curves. Eq. (\ref{eq:SR_anal2}) remains a tight upper bound for all scenarios but $U=2$. As anticipated in section \ref{subsub:unintended_sinr}, this is because of the correlation introduced in $\underline{W}$ which is large for small BOR values and which was neglected in the derivation of (\ref{eq:SR_anal2}). In addition, the SR obtained with the classical FD TR SISO OFDM system presented in Section \ref{subsec:traditional_FDTR}, i.e., no AN signal, is enhanced with the addition of AN except for very high percentages of AN. Furthermore, the SR increases when the BOR becomes higher because the TR gain becomes larger at Bob for higher BOR values but not at Eve. No more secrecy is obtained when $\alpha \to 0$, since the SINR's at Bob and Eve drop to zero.\\
Fig. \ref{fig:secrecy_alpha_bor_optimal} illustrates the values of $\alpha_{opt}$ given by (\ref{eq:best_alpha}) that maximize the SR determined from the closed-form approximation (\ref{eq:SR_anal2}), as well as obtained from the numerical simulations. The analytic estimation of the optimal amount of AN energy is not perfect but, the resulting simulated SR is very close to the maximal SR. The reason can be observed in Fig. \ref{fig:secrecy_alpha_bor} where the SR varies very slowly about its maximum when $\alpha$ changes. So, for a given BOR value, Alice can make a rough determination of $\alpha_{opt}$ and therefore the available SR, if $E_b/N_0$ is known.

\section{Conclusion}
In this paper, the problem of securing the FD TR SISO OFDM wireless transmission from a transmitter to a legitimate receiver in the presence of a passive eavesdropper is considered. A novel and original approach based on the addition of an AN signal onto OFDM blocks that improves the PLS is proposed. This approach can be easily integrated into existing standards based on OFDM. It only requires a single transmit antenna and is therefore well suited for devices with limited capabilities. Analytic and simulation results show that the novel approach significantly improves the security of the communication and so considerably jeopardizes any attempt of an eavesdropper to retrieve the data.
\label{sec:ccl}


\begin{thebibliography}{1}
 
 
 %
 
 
 
\bibitem{PLS_litt1} H. Alves, R. D. Souza, M. Debbah, and M. Bennis, ``Performance of transmit antenna selection physical layer security schemes'', in IEEE Signal Process. Lett., vol. 19, no. 6, pp. 372-375, 2012.


\bibitem{PLS_litt2} N. Yang, H. A. Suraweera, I. B. Collings, and C. Yuen, ``Physical layer security for TAS/MRC with antenna correlation'', in IEEE Trans. Inf. Forensics Security, vol. 8, no. 1, pp. 254-259, 2013.

\bibitem{PLS_litt3}D.-D. Tran, D.-B. Ha, V. Tran-Ha, and E.-K. Hong, ``Secrecy analysis with MRC/SC-based eavesdropper over heterogeneous channels'', in IETE Journal of Research, Mar. 2015. 
\bibitem{wyner} A. D. Wyner, ``The wire-tap channel'', in Bell Syst. Tech. J., vol. 54, pp. 1355-1387, Oct. 1975.


\bibitem{PLS_litt4} M. Li, S. Kundu, D.A. Pados, and S.N. Batalama, ``Waveform Design for Secure SISO Transmissions and Multicasting'', in IEEE Journal on Selected Areas in Communications, vol. 31, no. 9, Sep. 2013.

\bibitem{TR_FD_TD} T-H. Nguyen, J-F. Determe, M. Van Eeckhaute, J. Louveaux, P. De Doncker, and F. Horlin, ``Frequency-Domain Time-Reversal Precoding in Wideband MISO OFDM Communications Systems'', in arXiv e-prints, Apr. 2019.

\bibitem{TR_AN_2018_xu} Q. Xu, P. Ren, Q. Du, and L. Sun, ``Security-Aware Waveform and Artificial Noise Design for Time-Reversal-Based Transmission'', in IEEE Transactions on Vehicular Technology, vol. 67, no. 7, June 2018.

\bibitem{TR_AN_2018_Li} S. Li, N. Li, X, Tao, Z. Liu, H. Wang, and J. Xu, ``Artificial Noise Inserted Secure Communication in Time-Reversal Systems'', in IEEE Wireless Communications and Networking Conference, Apr. 2018.

\bibitem{TR_AN_2017_Li} S. Li, N. Li, Z. Liu, H. Wang, J. Xu, and X. Tao, ``Artificial Noise Aided Path Selection for Secure TR Communications'', in IEEE/CIC International Conference on Communications in China (ICCC), Oct. 2017.

\bibitem{otges} C. Oestges, A.D. Kim, G. Papanicolaou, and A. J. Paulraj, ``Characterization of Space-Time Focusing in Time-Reversed Random Fields'', in IEEE Transactions on Antennas and Propagation, vol. 53, no. 1, Jan. 2005.



\bibitem{TR_bor} T. Dubois, M. Crussi\`{e}re and M.   H\'{e}lard, ``On the use of Time Reversal for Digital Communications with Non-Impulsive Waveforms'', in 4th International Conference on Sig. Process. and Commun. Sys., Dec. 2010.
 

\bibitem{papr}  S. Ahmed, T. Noguchi, and M. Kawai, ``Selection of Spreading Codes for Reduced PAPR in MC-CDMA systems'', in  IEEE 18th International Symposium on Personal, Indoor and Mobile Radio Communications,  Sep. 2007 .

\bibitem{TR_Tran_secrecy_capa} H-V. Tran, H. Tran and G. Kaddoum
``Effective Secrecy-SINR Analysis of Time Reversal-Employed Systems over Correlated Multi-path Channel'', in IEEE $11^{\text{th}}$ international Conference on Wireless and Mobile Computing, Networking and Communications (WiMob), pp. 527-532, Oct. 2015

\bibitem{TR_la} M. M. Molu, P. Xiao, M. Khalily, L. Zhang, and R. Tafazolli, ``A novel equivalent definition of modified Bessel functions for performance
analysis of multi-hop wireless communication systems,'' in IEEE Access,
vol. 5, pp. 7594–7605, May 2017.


\end{thebibliography}
\end{document}